\documentclass[11pt]{article}
\usepackage{times}
\usepackage{geometry}
\usepackage[utf8]{inputenc}
\usepackage{enumitem,amssymb}
\usepackage{graphicx}
\usepackage{fancyhdr}
\usepackage{aas_macros}
\usepackage{mdframed} 
\usepackage{wrapfig}

\usepackage[authoryear]{natbib}
\setcitestyle{authoryear,open={(},close={)}}

\mdfdefinestyle{theoremstyle}{
innertopmargin=\topskip,}
\mdtheorem[style=theoremstyle]{lrptextbox}{}

\usepackage[symbol]{footmisc}

\begin{document}
\begin{center}

{\Large Signposts of planet formation in protoplanetary disks}\\
\vspace{4mm}


Nienke van der Marel$^{1}$\footnote[1]{astro@nienkevandermarel.com}, Ruobing Dong$^2$, Ralph Pudritz$^3$, James Wadsley$^3$, Aaron Boley$^4$, Eve Lee$^5$, \\ Mohamad Ali-Dib$^6$, Brenda Matthews$^{1,2}$, Christian Marois$^{1,2}$, Henry Ngo$^1$\\
\emph{\small 1) NRC Herzberg, 2) University of Victoria, 3) McMaster University, 4) University of British Columbia,\\ 5) McGill University, 6) University of Montreal}\\
\end{center}








\begin{abstract}
Successful exoplanet surveys in the last decade have revealed that planets are ubiquitous throughout the Milky Way, and show a large diversity in mass, location and composition. At the same time, new facilities such as the Atacama Large Millimeter/submillimeter Array (ALMA) and optical/infrared facilities including Gemini/GPI have provided us with sharper images than ever before of protoplanetary disks around young stars, the birth cradles of planets. The high spatial resolution has revealed astonishing structures in disks, such as rings, gaps, asymmetries and spiral arms, and the enormous jump in sensitivity has provided the tools for both large, statistically relevant surveys and deep, sensitive molecular line studies. These observations have revolutionized our view of planet formation, disk formation and disk evolution, bringing model simulations and observations closer to the same level of detail, with many contributions from Canadian researchers on theoretical, observational and technological sides. The new results have inevitably led to a range of new questions, which require next generation instruments such as the Next Generation Very Large Array (ngVLA) and large scale optical infrared facilities. In this white paper we will discuss the current transformation in our understanding of planet formation and the next steps and challenges in connecting theory with exoplanet demographics and protoplanetary disk observations for Canadian research. 
\end{abstract}

\newpage
\section{Introduction}
\begin{figure}[!ht]
\centering
\includegraphics[width=0.8\textwidth,trim=0 0 0 0]{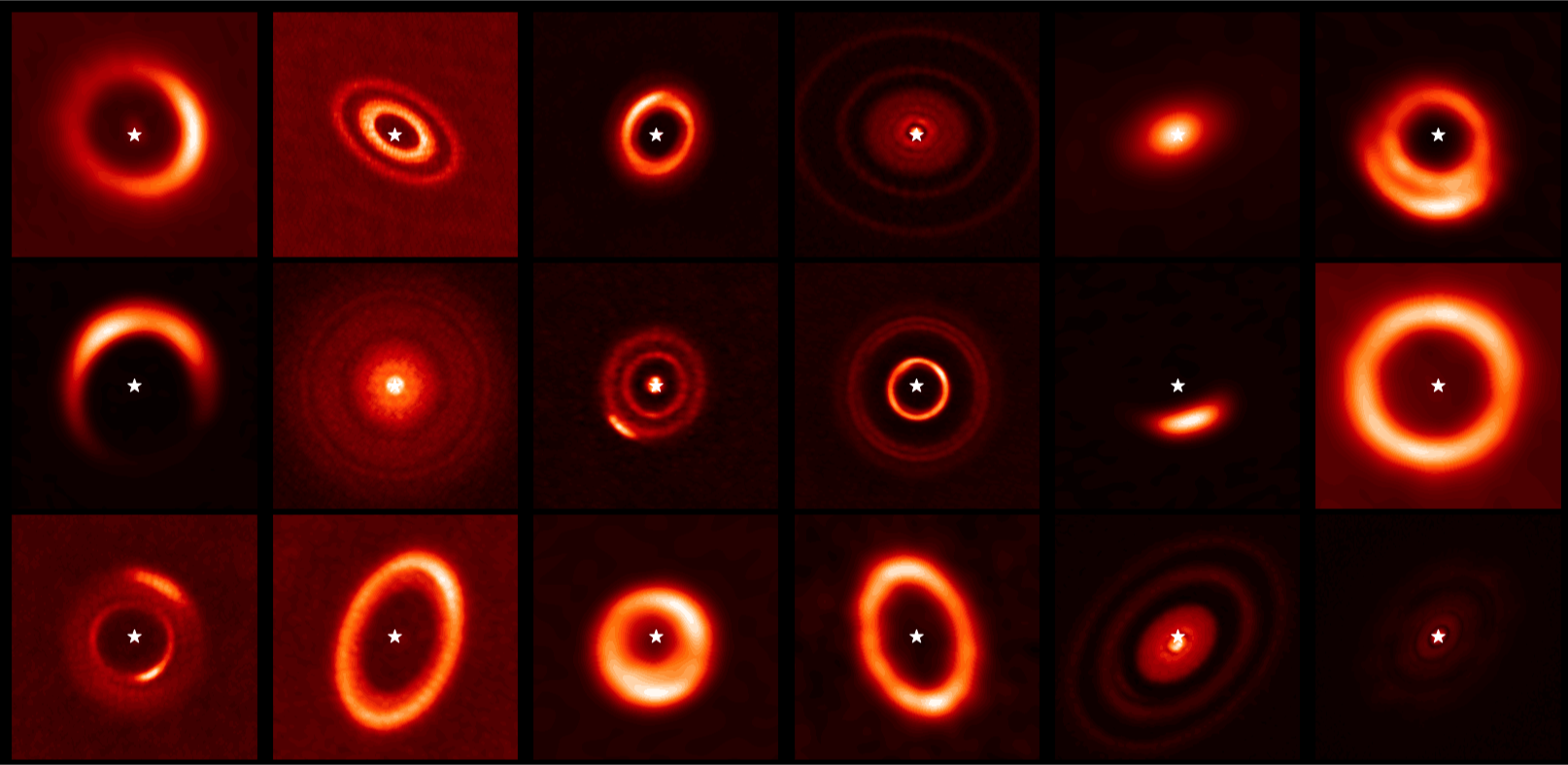}
\caption{\emph{\small Overview of disks observed with ALMA (credits Nienke van der Marel). There is clearly a large diversity of structures, including gaps, rings, asymmetries and clumps, which are all thought to be related to the dust trapping phenomenon.}} 
\label{fig:overview}
\end{figure}

\begin{wrapfigure}{r}{0.5\textwidth}
\includegraphics[width=0.5\textwidth,trim=0 20 20 0]{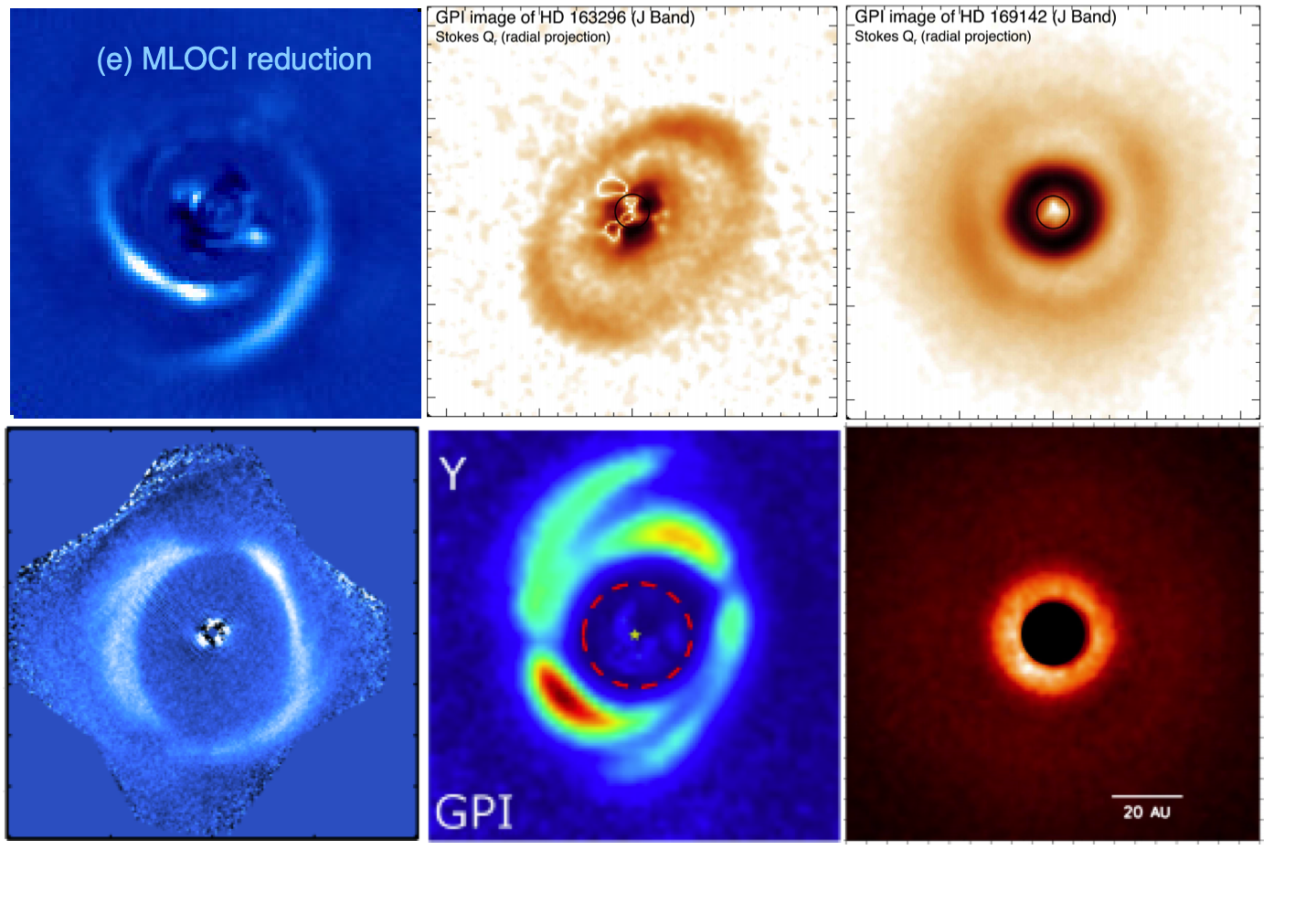}
\caption{\emph{\small Overview of disks observed with Gemini/GPI in scattered light. From left to right, bottom to top: HD135344B \citep{Wahhaj2015}, HD163296, HD169142 \citep[both from][]{Monnier2017}, HD142527 \citep{Rodigas2014}, HD100453 \citep{Long2017} and TW Hya (Rapson et al. 2015). In scattered light disks show rings, shadows and spiral arms, linked to the presence of planets in the disk. }}
\label{fig:gpi}
\end{wrapfigure}

\vspace{-2mm}
With the discoveries of thousands of exoplanets, our understanding of the formation of our Solar System has been transformed into a much more complex question: how can we understand the large diversity of planets in size, location and composition throughout the Galaxy? We tackle these problems by going back to the beginning: the protoplanetary disk, in which these planets are being born. Protoplanetary disks of gas and dust are found around young stars of $\lesssim$10 Myr old that are considered the natural consequence of star formation (E025: DiFrancesco): during the gravitational collapse of a molecular cloud the conservation of angular momentum leads to the formation of a flattened, rotating structure around the newly formed star, called a `disk'. This disk continues to accrete material from the cloud. Eventually the gas dissipates, giving rise to a collisionally dominated system, called a debris disk, which contains planets, planetesimals, and dust (E016: Matthews). 
The interactions between planets, the star, and the protoplanetary disk are imprinted onto readily observed disk structures, in particular observed in the last decade by the Atacama Large Millimeter/submillimeter Array or ALMA (Figure \ref{fig:overview}) and NIR imaging instruments such as Gemini/GPI, VLT/SPHERE and Subaru/HiCiAO/SCExAO (Figure \ref{fig:gpi}). Coupled with theoretical disk simulations and our knowledge of stellar populations and the Solar System, studying protoplanetary disks can chart the early stages of planet formation, identifying the birth sites, masses, and atmospheric compositions of exoplanets.

Until 2010, protoplanetary disks were primarily identified through Spectral Energy Distributions (SEDs) of young stars, as they generally exhibit a large excess in infrared and millimeter wavelengths above the stellar photosphere, originating from the thermal dust emission in the disk. \emph{Spitzer} (2003-2009), \emph{Herschel} (2009-2013) and JCMT (1987-), all infrared/submillimeter telescopes with  Canadian involvement, have greatly contributed to the search for and have found hundreds of protoplanetary disks in nearby star forming regions \citep[$\sim$100-200 pc, e.g.][]{Evans2009}. Knowledge of the structure of disks was limited to the brightest, largest ones through interferometric (SMA) observations showing dust grains out to $\sim$100 au and a gas disk out to several 100s of au. A fraction of disks was found to be hosting large inner dust cavities, the so-called \emph{transition disks} \citep{Andrews2011}. 

Modern resolved observations of protoplanetary disks with angular resolutions on the order of tens of mas have truly revolutionized the view of protoplanetary disks. On one hand, numerous discoveries have been made since the commission of ALMA in 2011. 
ALMA allows, for the first time, fully resolved observations of disks in both dust and gas at spatial scales of a few au, similar to the orbits of the giant planets in our Solar System, and detections of disks down to a fraction of an Earth mass in dust in only a few minutes of observing time. Not surprisingly, the top 3 most-cited ALMA publications (out of over 1500) report on observations of protoplanetary disks: the iconic HL~Tau paper \citep{HLTau2015}, together with the TW Hya disk \citep{Andrews2016}, revealed multiple possibly planet-induced gaps in these planet cradles;
and the first paper demonstrating the presence of dust trapping in the Oph~IRS~48 system, a phenomenon required for efficient dust growth beyond millimeter sizes and to trigger planetesimal formation \citep{vanderMarel2013}.

Transformational improvements have also been made in the last decade in the area of near-IR imaging (e.g.
 Gemini). Adaptive optics reaching sub-arcsecond resolution, in combination with coronographs with inner working angles of $\sim$0.1", have enabled imaging in scattered light down to $\sim$15 au separation from the star, and searches for companions in the inner part of the disk. NIR imaging has led to the discoveries of numerous disks with spiral arms and dust gaps (see Figure \ref{fig:gpi}), both signposts of planets.

Canada has played an important role in the recent advances at both wavelengths. ALMA was identified as `Canada's highest priority for participation in a major ground-based observatory' in the first Canadian LRP . Canada has been involved in e.g. the design and construction of the Band 3 receivers, software development, and representation in e.g. the ALMA Board and ALMA Science Advisory Committee. Furthermore, Canada is expected to play a major role in planned future interferometer arrays such as ngVLA and SKA, natural complementary facilities to study protoplanetary disks at even longer wavelengths. Also optical/infrared facilities such as Gemini, JWST and ELTs (in which Canada is partner) will answer key questions in protoplanetary disk studies in the coming decade.\ \emph{Considering the strong technical involvement of Canada in these observatories, combined with the powerful Canadian computational facilities, the current scientific expertise in star formation, exoplanets, debris disks and planet formation theory, expansion in the fast-growing research field of observational planet formation is naturally encouraged, in order to take maximum advantage of the Canadian investments.}

In the past decade, the field of planet formation and protoplanetary disks has been transformed through the arrival of ALMA and other telescopes.
Naturally, observational results have also led to new mysteries. We will evaluate the current status of the field through the discussion of three major questions in the following section:
\begin{enumerate}
\setlength{\itemsep}{0pt}
\item How do disks evolve after formation?
\item How and where do planets form and how does disk evolution affect planet formation?
\item How does the disk affect the properties and compositions seen in exoplanets?
\end{enumerate}
We will further discuss future directions and possibilities for the Canadian astronomical community.

\vspace{-2mm}
\section{Current results of protoplanetary disks}
\label{sect:current}
\vspace{-2mm}
\subsection{How do disks evolve after formation?}
\label{sect:df}
\vspace{-2mm}
One of the major questions in planet formation is how gas-rich, optically thick protoplanetary disks evolve and dissipate. 
Traditionally, this process is described by the viscous $\alpha$-disk model \citep{ShakuraSunyaev1973}, in which accretion driven by turbulence accompanying outward transporting angular momentum leads to a gradual spreading of the disk during its lifetime \citep{Hartmann1998}. Turbulence was thought to be provided through the magneto-rotational instability or MRI \citep[e.g.][]{Balbus1991}

Eventually stellar photoevaporation dissipates the disk when the accretion has dropped below a critical threshold. Whereas this scenario appeared to be supported by the typical lifespan of disks of a few Myr  
\citep[e.g.][]{Fedele2010}, and a correlation between disk mass and accretion rate \citep{Manara2016}, recently measured weak turbulence in disks \citep{Flaherty2018} and the (re-)evaluation of non-ideal MHD effects led to the conclusion that MRI is inefficient in providing turbulence and disk accretion may be driven by (magnetic) disk winds \citep{Bai2013,Pudritz2019} or planetary torques \citep[e.g.][]{FungChiang2017}. These scenarios cannot be observationally constrained yet with current facilities. 

ALMA snapshot disk surveys of nearby star forming regions have revealed that the disk dust mass decreases with time \citep[e.g.][]{Ansdell2018}.
On the other hand, it is clear that the disk dust masses of typical protoplanetary disks are well below the threshold required to produce the Solar System \citep{Hayashi1981} or exoplanet populations \citep{Winn2015}. Either the estimates of the dust masses are too low (due to uncertainties in dust opacity, dust size and optical depth), 
or planet formation occurs on rapid timescales during the embedded disk phase in $<$1 Myr \citep{Najita2014,Manara2018} 


How protoplanetary disks evolve into debris disks remains unclear as well. 
Transition disks with large inner cavities are thought to be cleared by giant planets \citep[see][for a review]{vanderMarel2017}. For a long time, they were thought to be an evolutionary class of disks \citep{Strom1989}.
However, transition disks with $>$20 au cavities appear to be found only in higher-mass disks \citep{OwenClarke2012}, and the occurrence rate of distant giant planets is too low 
for them to be a common phase that all disks go through \citep{Dong2016,vanderMarel2018}. Debris disks are also fundamentally different from protoplanetary disks, as their dust and gas is considered to be of secondary origin: the result of planetesimal collisions. A more likely disk class connecting these two populations is the \emph{hybrid disk}, a dust debris disk with evidence for primordial gas 
, such as the 5 Myr old HD141569 \citep{White2016,White2018}. Only a handful of such systems are known, suggesting that the gas dispersal phase is likely rapid in most systems.



\vspace{-2mm}
\subsection{How and where do planets form and how does disk evolution affect planet formation?}
\label{sect:pf}

\begin{figure}[!ht]
\includegraphics[width=0.75\textwidth]{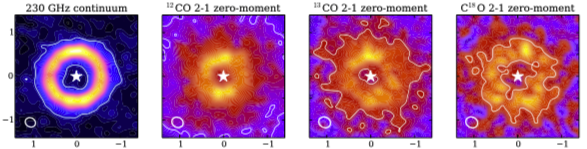}
\includegraphics[width=0.2\textwidth]{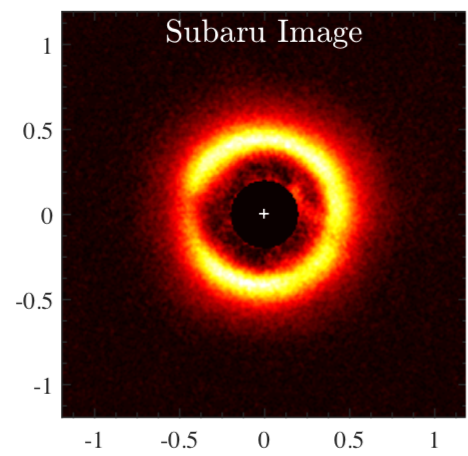}
\caption{\emph{\small ALMA observations of the J1604-2130 disk of the dust (left) and gas (through three CO isotopologues, $^{12}$CO, $^{13}$CO and C$^{18}$O) and the scattered light image (right) \citep{Dong2017}. The deep gap indicates the presence of a giant planet.}}
\label{fig:j1604}
\end{figure}

\vspace{-2mm}
The formation of rocky planets and the cores of giant planets is generally described as follows: ISM micrometer-sized dust grains coagulate to planetesimals, which subsequently which subsequently grow even larger by accreting smaller pebbles and/or collision with other planetesimals. Next, they accrete gas from the disk at a rate regulated by internal cooling, followed by rapid runaway growth in response to gas self-gravity \citep{Pollack1996}. The primary factors in this process related to protoplanetary disk studies are the disk mass, governing the dynamics and amount of available material; the timescale of disk dissipation, setting the timescale of rocky core formation; and the growth of small dust grains to centimeter-sized pebbles, which requires overcoming a number of barriers \citep{Birnstiel2010}. Beyond pebble sizes dust growth can no longer be observationally constrained and studies are  theoretical only: pebbles are thought to grow into planetesimals and rocky cores through streaming instability \citep[e.g.][]{Youdin2005} and subsequently pebble accretion \citep[e.g.][]{Lambrechts2012}.

Recent progresses in this direction, in which Canadian researchers have made major contributions, are on theoretical models of planet formation, with a focus on explaining exoplanet population diversity: 
formation of solid cores of gas giants through pebble accretion and planetesimal scattering \citep{Levison2015}; 
connecting disk dust properties with the streaming instability (Rucska \& Wadsley, in prep.);
the onset and timing of gas accretion onto rocky cores \citep{LeeChiang2016,Lee2019,FungLee2018};
the evaporation valley or Fulton-gap of close-in planets of radii between 1.5 and 2.0 $R_{\rm Earth}$ \citep{OwenWu2013,OwenWu2017};
constraining the brightness of gas giants through cooling curves \citep{marleau},
the presence of hot Jupiters without the need for migration \citep{Boley2016};
and the dynamical interactions between tightly-packed inner planets (STIPs) and a planet at larger orbital distances \citep{Granados2018}. 
Many of these studies are ultimately linked to our understanding of disk mass, evolution and dissipation, which remain highly uncertain as discussed in the previous section.

For the first step of dust growth, ALMA observations have provided firm evidence for the existence of dust trapping in pressure bumps. Multi-wavelength observations show that millimeter-sized grains are more concentrated than the gas and micrometer-sized grains in disks \citep[e.g.][]{vanderMarel2013,vanderMarel2015-12co}, providing a solution for the historic radial drift problem preventing growth beyond millimeter sizes \citep{Weidenschilling1977}. Rings, asymmetries, and spiral arms in continuum emission of millimeter grains are now commonly observed \citep{vanderMarel2019} and e.g., the DSHARP ALMA Large Program \citep{Andrews2018dsharp},
and generally interpreted as dust traps \citep[e.g.][]{ Huang2018}. However, the quantification of dust growth through millimeter observations remains limited as the dust emission is often optically thick.

An alternative to the core accretion model is the possibility that giant planet formation occurs rapidly through gravitational instability and disk fragmentation \citep{Boss1997}. This scenario, in particular the cooling mechanism, has been explored by Canadian researchers as well \citep{Rogers2011,Rogers2012,Boley2016}. This formation scenario has been out of favor considering the low disk masses and surface densities measured in disks, but considering the large uncertainties it can certainly not be excluded \citep{Dong2018spirals}. In particular, if planet formation happens in the embedded stage, where disks are more massive and compact, gravitational instability may play a key role in forming giant planets early on \citep{Boley2009}.

\begin{wrapfigure}{r}{0.5\textwidth}
\includegraphics[width=0.5\textwidth,trim=0 20 20 0]{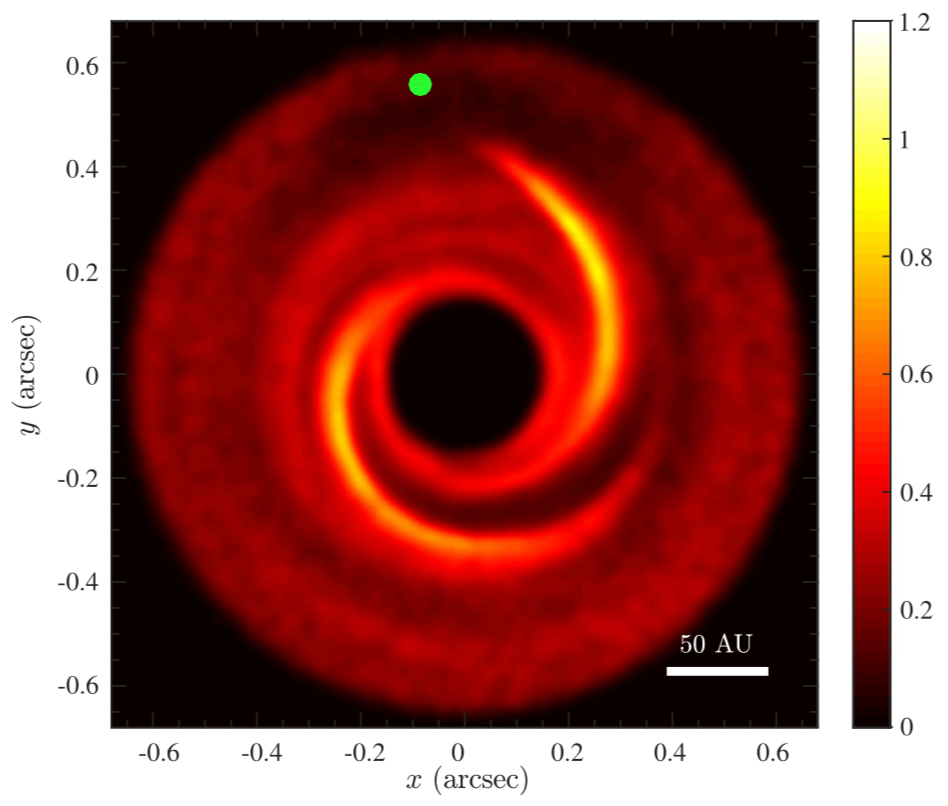}
\caption{\emph{\small Result of a simulation of planet-disk interaction with a planet in the outer part of the disk, which triggers spiral density waves which are thought to be observed in scattered light images with e.g. Gemini/GPI (Figure \ref{fig:gpi}). Figure taken from \citet{Dong2015spirals}. }}
\label{fig:spirals}
\end{wrapfigure}
Rings and asymmetries are interesting sites of study of dust evolution and dust trapping, but also the physical processes \emph{creating} these structures are relevant for our understanding of planet formation. The origin of the millimeter-dust gaps remains inconclusive: the large inner cavities in transitional disks appear to be consistent with being opened by distant giant planets ($>$10 au) \citep{Crida2006,Fung2014,Dong2015} considering the presence of gas cavities and scattered light cavities inside the mm-dust cavities \citep[e.g.][]{Dong2012, vanderMarel2013,vanderMarel2016-isot,Dong2017}.
However, for the narrow gaps in the outer disk the gas structure remains unconstrained \citep{Isella2016,vanderMarel2018-hd16}. The narrow width of the dust gaps in the DSHARP survey suggest Saturn and Neptune-mass planets to be responsible \citep{Zhang2018dsharp}, as planets with masses as low as Super-Earths can open dust gaps in low-viscosity disks \citep{Dong2017multi,Dong2018gaps}. 
Scattered light images of e.g. Gemini/GPI show rings and gaps in numerous disks as well \citep[e.g.][and Figure \ref{fig:gpi}]{Monnier2017}.

Disk observations thus indicate that embedded planets must already be present. In this early stage the common way to detect planets is direct imaging as traditional methods such as transits and radial velocity are hindered by the intrinsic stellar variability in pre-main sequence stars. Direct imaging is a technique well established in Canadian research, as evidenced by the discovery of the HR8799 multiple-planet system \citep{Marois2008}. However, direct imaging searches for planets in protoplanetary disk with instruments like Gemini/GPI and VLT/SPHERE have so far resulted in only one firm detection in the PDS~70 disk \citep{Keppler2018} and numerous non-detections. Surveys of giant planets at wide orbits (10--100 au) around main-sequence stars such as the Gemini Planet Imager Exoplanet Survey (GPIES) suggest that such planets are rare under the assumption that they are formed through the hot-start model \citep[e.g.][]{Nielsen2019} and cannot explain the high frequency of disks with gaps at large radii. On the other hand, constraints on planets responsible for the narrow gaps in DSHARP may simply be below the detection limits of current instruments \citep{Zhang2018dsharp}. 
Finally, the planet scenario described above creates a chicken-and-egg problem: if dust traps are created at planet gap edges, how are the first planets formed if trapping is required to enhance dust growth?

Other explanations for the presence of gaps in disks include snowlines \citep{Zhang2015}, 
planetesimal collisions \citep{Boley2017}, and secular gravitational instabilities \citep{Takahashi2016}. The scenario involving snowlines (disk radii where molecules freeze out and increase the local dust growth) predicts a clear correlation of gap locations with disk temperature, which is not observed in large gap samples
\citep{Huang2018, vanderMarel2019}. On the other hand, time-dependent disk models including viscous evolution of the disk predict that the H$_2$O ice line moves inwards with time \citep{Cridland2017}; as disks age and viscous evolution remains highly uncertain, the snowline scenario may be applicable in certain cases. The other scenarios are difficult to test observationally and the planet scenario remains the most popular.

The presence of planets in disks is also implied by the detections of other features in disks, in particular spiral arms, that are thought to be planet-excited density waves \citep{KleyNelson2012,Dong2015spirals} and detected in scattered light imaging with Gemini/GPI, VLT/SPHERE and Subaru/HiCiAO \citep[e.g.][]{Muto2012,Wahhaj2015,Follette2017}. The morphology of the spiral arms can be used to derive properties of the (undetected) planets \citep[][and Figure \ref{fig:spirals}]{Fung2015, Dong2017spirals}. Misalignments between the inner and outer disk have been observed through shadows and non-Keplerian motion \citep[e.g.][]{Marino2015}, which can be readily explained by the presence of a massive companion warping and breaking the disk \citep[e.g.][]{Lodato2013}. The lack of detections of companions in disks remains an issue in testing and quantifying these signatures though.



\vspace{-2mm}
\subsection{How are protoplanetary disks connected to exoplanet properties?}
\label{sect:ep}
\vspace{-2mm}
Protoplanetary disks are expected to leave imprints onto the planets formed in them. Exoplanet properties such as mass, composition, and orbital radius can be compared with formation models, although one has to take into account significant changes in these properties between formation and final status. 

An important aspect in this context is the final location of the planet. In the early stages of planet formation, physical processes are dominated by the gas dynamics in the disk. Gas is needed to form gas giants, but it also causes rapid ($\sim10^5$ yrs) inward planetary migration \citep{Ward1997}. Both inviscid disks \citep{FungChiang2017} and \emph{planet traps} \citep[disk inhomogeneities in density, temperature or viscosity, see][]{Hasegawa2011} have been proposed to halt migration. Three main planet traps have been explored in a continuous study by  Canadian researchers: the heat transition,
the snowline,
and the dead zone.
Planet traps move inwards with time as the accretion rate drops, even more significantly with the inclusion of dust evolution \citep{Cridland2017}. Time-dependent disk models can readily predict exoplanet populations in the mass-period diagram \citep{Hasegawa2013} and the range of Super Earth compositions \citep{Alessi2017}. Also snowlines have been investigated as preferential locations for planet formation \citep{alidib2}.

One of the hot topics in Canadian astronomy is the characterization of exoplanetary atmospheres, as they form a direct link to the possibility of detection of life outside our Solar System \citep[see][for a review]{Cowan2015}. Molecules detected in exoplanetary atmospheres include e.g. H$_2$O, CO$_2$, CO, CH$_4$ and HCN. The C/O ratio can be used as a tracer of the location where the planet is formed, as the C/O ratio varies with radius throughout the disk due to the ice line of volatiles like H$_2$O, CO and CO$_2$ \citep{Oberg2011snowline}, as demonstrated for e.g. hot Jupiters  \citep{alidib}. Observed ratios of exoplanet atmosphere constituents match well with the predictions of the time-dependent planet trap disk models coupled with dust and chemical evolution, and nitrogen carrying molecules such as NH$_3$ can be used to inform the planet formation history of a system \citep{Cridland2016,Cridland2017chemistry}. 

Deriving the planet formation history in this way requires a solid knowledge of the disk chemistry. ALMA has detected numerous molecules in disks, including CO, CN, CS, H$_2$CO, CH$_3$OH, C$_3$H$_2$, CH$_3$CN, C$_2$H and N$_2$H+ \citep[e.g.][]{Oberg2015}, which help to constrain chemical disk models \citep[e.g.][]{Furuya2014} and can be linked with the chemical complexity in star forming regions and comets in the Solar System \citep{Herbst2009}. The molecular abundances are sparsely constrained and disk chemistry is far from understood. 

Even our knowledge of the chemistry of CO in disks is likely incomplete: CO isotopologue data are used to estimate disk gas masses using chemical models \citep{WilliamsBest2014,Miotello2016} but the average masses are well below the Solar System nebula, implying planet formation must happen in the first Myr in the embedded stage; or the disk mass is underestimated due to additional chemical effects leading to carbon depletion, where carbon is locked up in grains and/or complex molecules \citep{Miotello2017}. The consequences of the uncertainty in disk gas mass and gas-to-dust ratio is severe for every aspect of planet formation studies. 

\vspace{-2mm}
\section{2020-2030: new opportunities for studies of planet formation}
\vspace{-2mm}
The future is bright for protoplanetary disk research. With partnership in new astronomical facilities such as the JWST, Gemini, thirty-meter-class telescopes, ALMA, EVLA/ngVLA and SKA, Canada is in an excellent position to invest in observational planet formation research. This will help constrain theoretical models of planet formation by better knowledge of disk mass, composition, evolution and dissipation timescales. The numerous discoveries with ALMA and other facilities, and the new insights from theoretical models, have led to a large number of important questions that will be accessible in the next decade, moving the field of planet formation forward by making direct connections with exoplanet properties. As high resolution observations with ALMA, GPI, and others continue, we propose a number of opportunities with new and planned facilities for Canada:

\begin{figure}[!ht]
\centering
\includegraphics[width=0.8\textwidth]{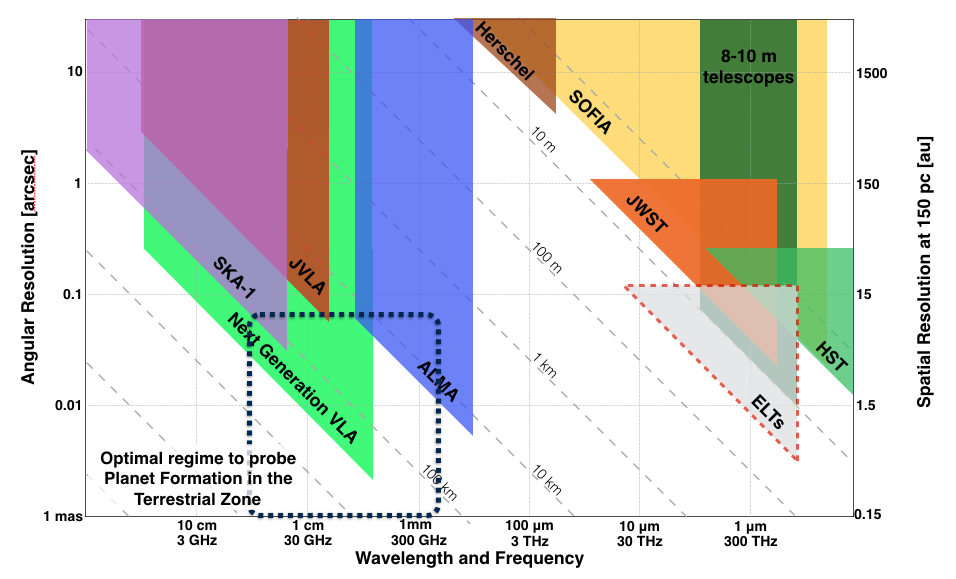}
\caption{\emph{\small Range of frequency and spatial resolution as traced by various facilities. The primary regime for studying planet formation in the terrestrial regime is indicated by the black dashed square, making the ngVLA the preferred facility for further study of planet formation. SKA will provide useful information as well on larger scales. Figure courtesy by Andrea Isella. }}
\label{fig:ngvla}
\end{figure}

\begin{itemize}
\setlength{\itemsep}{0pt}
\item The study of dust growth and measurement of dust masses can be greatly improved by going to centimeter wavelengths at subarcsecond resolution, with the addition of the Band 1 and 2 receivers (35-52 and 67-90 GHz, respectively) to ALMA and with the planned ngVLA and SKA facilities in the centimeter regime. As millimeter continuum emission is often optically thick, substructures in ALMA images may remain hidden in the inner $\sim$5 au of the disk, which is the most interesting region of interest for terrestrial planet formation (see \ref{sect:pf}). Dust growth in dust traps up to centimeter grains is hinted at in VLA observations of a handful of disks \citep[e.g.][]{vanderMarel2015vla}, but higher angular resolution of $\sim$0.01" ($\sim$1-2 au at the typical distance of nearby disks) will truly help to constrain dust growth in the optically thin regime (see \ref{sect:pf}). Centimeter disk surveys will allow a proper measurement of disk dust masses and assess the potential of disks to form planets and the disk dissipation process (see \ref{sect:df}).

\item High angular resolution centimeter observations allow to explore a whole new area in protoplanetary disks: free-free emission can originate from ionized gas in disk winds, which are predicted as an alternative for the viscous disk model to carry away angular momentum.  The viscous disk model is currently under debate (see \ref{sect:df}) with important consequences for disk evolution and planet formation time scales, but in lack of proof of existence of disk winds the disk evolution process remains unknown. SKA and ngVLA will provide the means to detect and quantify these winds in free-free emission \citep{Pascucci2012} and revolutionize our understanding of disk evolution.

\item A firm connection between the observed disk structures and embedded young planets requires the detection of planets in the disk itself. Further upgrades to the Gemini/GPI and its AO system and the future opportunities of an ELT will allow to detect planets at much higher contrast and at closer-in angles, tracing into the regime of Neptune to Saturn mass planets at tens of au \citep{Dong2017gaps}, predicted by the disk structures as observed by ALMA. 

\item An indirect way of detecting embedded planets in disks is through their circumplanetary disk: ALMA (and in the future, ngVLA) is able to detect circumplanetary disks, as demonstrated by two recent detections \citep{Isella2019,PerezS2019} and kinematic signatures of planets in disks \citep{Pinte2019}. More discoveries are expected as this possibility has been demonstrated for the brightest disks only. Characterization of these circumplanetary disks will provide further insight on the planet accretion process.

\item Further theoretical modeling of planet formation, simulations of planet-disk interaction (in particular predicting and quantifying signatures of embedded planets, e.g. spirals and gaps), and migration vs planet traps, which requires further expansion of the Compute Canada network. Theoretical modelers need to work side by side with observational experts, for a proper interpretation of both data and theory.

\item Exoplanet atmospheric characterization with the JWST and ELTs will provide further means for our understanding of their formation history in the disk. Knowledge of the composition of exoplanets in the habitable zone is also interesting for studies of astrobiology and the origin of life. Disk chemistry will be further constrained with observations of complex organic molecules, which are generally produced in ices and non-thermally desorbed at low temperatures in the centimeter traced by SKA and ngVLA. Inner disk chemistry, in particular the H$_2$O snowline, can be studied with JWST and ELTs in the near infrared, tracing the terrestrial regime inaccessible by ALMA.
\end{itemize}



\vspace{-2mm}
\section{Requirements}
\vspace{-2mm}
Realizing the goals for the coming decade and expand Canada's role in the exciting area of planet formation research will require:
\begin{itemize}
\setlength{\itemsep}{0pt}
\item Hiring experts on observational planet formation, on faculty, postdoc and graduate level.
\item Better integration of related research areas such as star formation, exoplanets, debris disks, Solar system, dynamics and theoretical disk studies throughout the country.
\item More training in use of interferometric facilities such as ALMA and EVLA, in particular for graduate students, and training to preparer for JWST and ELTs.
\item Major collaborations of planet formation experts, within and outside of Canada, for a full interpretation of the complex problems in planet formation studies. 
\item Investing and partnership in future facilities for centimeter observations such as ngVLA and SKA. For protoplanetary disk studies the ngVLA is preferred considering its higher frequency range and resolution, which allows us to trace the disk structure in the terrestrial regime. However, SKA will provide useful information as well on larger scales. 
\item Further support for computational facilities, in particular Compute Canada, both for data reduction and theoretical modeling.
\item Continued access to and upgrades of Gemini/GPI for improved capabilities of direct imaging of exoplanets in disks, in particular inner working angle and use of fainter guide stars, until equivalent/better options at an ELT are avaiable.
\end{itemize}

\vspace{-2mm}
\section{Canada's strengths}
\label{sect:strengths}
\vspace{-2mm}
Observational planet formation is currently a compact community in Canada due to its novelty. However, world-wide this is one of the fastest growing fields in astronomy, due to huge advancements in the last decade in observations. Canada has all the ingredients to be one of the major contributors to this exciting field: 

\begin{itemize}
\setlength{\itemsep}{0pt}
\item A long tradition of star formation and exoplanet research.
\item Numerous theoretical studies linking exoplanet properties to disks (see Section \ref{sect:current}), providing important contributions to the biggest questions in planet formation research.
\item Excellent technological expertise in instrument technology in both millimeter and optical wavelengths for ALMA, EVLA, Gemini, JWST and ELT instruments.
\item ALMA data reduction expertise with HAA as one of the few ALMA ARC nodes in the world 
\item Exceptional computational and data archival facilities through ComputeCanada, CANFAR and CADC, crucial for processing and storing of large amounts of data generated by  current and next generation telescopes. 
\item A unique data processing platform for ALMA data, called ARCADE (ALMA Reduction in the CANFAR Data Environment), which recently received NRAO funding for a full development study to use ARCADE as a public facility (PI Helen Kirk). ARCADE is a web-accessible, fully interactive computing environment tailored to ALMA users.
\end{itemize}

\vspace{-2mm}
\section{Long term perspective}
\vspace{-2mm}
ALMA, ngVLA and SKA will continue to deliver exceptional millimeter and centimeter images of disks for the next few decades. The current 10m-class telescopes such as Gemini may be superseded by 30m-class telescopes in less than a decade. Further plans for space observatories, in particular in the mid- and far infrared, will further enhance our knowledge of protoplanetary disks and planet formation. Infrared interferometric facilities will be of particular interest as they will provide the means to zoom into the habitable zones where terrestrial planets form.

\vspace{-2mm}
\section{LRP criteria}

\begin{lrptextbox}[How does the proposed initiative result in fundamental or transformational advances in our understanding of the Universe?]
\vspace{-2mm}
How does the solar system, and exoplanetary systems, form are some of the most fundamental quests of humankind in the search for our origin and our place in the Universe. With observing facilities in the last decade enabling direct observations of planet formation for the first time in history, we are at the brink of making transformational advances in this area.
The next steps in observational planet formation are 1) disk mass; 2) disk dissipation mechanisms and disk wind; 3) signatures of embedded planets setting constraints on planet formation timescale; and
4) exoplanet atmosphere composition. Advanced in these areas will make major impacts to our understanding of how planets and the solar system form, and of the diversity of exoplanets.
\end{lrptextbox}

\begin{lrptextbox}[What are the main scientific risks and how will they be mitigated?]
\vspace{-2mm}
Part of this research depends on the realization of future facilities such as ngVLA and SKA. However, even without those, a large amount of research can be performed with existing and approved facilities (and upgrades) such as ALMA, Gemini, JWST and ELTs.
\end{lrptextbox}

\begin{lrptextbox}[Is there the expectation of and capacity for Canadian scientific, technical or strategic leadership?]
\vspace{-2mm}
Yes. As explained in Section~\ref{sect:strengths}, Canada has accesses to the few most important observing instruments in the field, now and in the future. The community is also well supported by excellent computing and data management infrastructures.
\end{lrptextbox}

\begin{lrptextbox}[Is there support from, involvement from, and coordination within the relevant Canadian community and more broadly?] 
\vspace{-2mm}
Yes. The Canadian community on planet formation is closely involved in this exciting new field and is in coordination to support the effort, as demonstrated by this white paper. However, due to the novelty of the field, the community is still in its development phase. Further support from the LRP and Canadian astronomy is needed to fully exploit the expertise in our community and Canada's accesses to observing facilities.
\end{lrptextbox}

\begin{lrptextbox}[Will this program position Canadian astronomy for future opportunities and returns in 2020-2030 or beyond 2030?] 
\vspace{-2mm}
Yes. Investment in planet formation and the suggested facilities (e.g., ngVLA) will position Canadian astronomy for many opportunities in planet formation research both on short term (2020-2030) in use of ALMA/ Gemini/JWST facilities and development of future facilities such as ngVLA/SKA, which are expected to deliver scientific results in the 2030s.
\end{lrptextbox}

\begin{lrptextbox}[In what ways is the cost-benefit ratio, including existing investments and future operating costs, favourable?] 
\vspace{-2mm}
N/A
\end{lrptextbox}

\begin{lrptextbox}[What are the main programmatic risks
and how will they be mitigated?] 
\vspace{-2mm}
N/A
\end{lrptextbox}

\begin{lrptextbox}[Does the proposed initiative offer specific tangible benefits to Canadians, including but not limited to interdisciplinary research, industry opportunities, HQP training,
EDI,
outreach or education?] 
\vspace{-2mm}
Planet formation is a broad field of research, with a strong connection to questions about origin of life and astrobiology: these topics together with the astonishing disk images that have been delivered and are expected provides many opportunities for outreach and education. Worldwide this field attracts many students because of its exciting observational discoveries. The proposed initiative will make Canada competitive in the global competition of talents.
\end{lrptextbox}

\twocolumn


\begin{thebibliography}{}
\setlength{\itemsep}{0pt}\small
\bibitem[\protect\citeauthoryear{{Alessi}, {Pudritz}, \& {Cridland}}{{Alessi}
  et~al.}{2017}]{Alessi2017}
{Alessi*}\footnotetext{*) References with major Canadian contributions.}, M., et~al. 2017, \mnras, 464, 428

\bibitem[\protect\citeauthoryear{Ali-Dib}{2017}]{alidib} Ali-Dib*,  M., 2017, MNRAS, 467, 2845

\bibitem[\protect\citeauthoryear{Ali-Dib et al. }{2017}]{alidib2} Ali-Dib*, M., et al. 2017, MNRAS, 469, 5016

\bibitem[\protect\citeauthoryear{{ALMA Partnership} et~al.}{{ALMA Partnership}  et~al.}{2015}]{HLTau2015}{ALMA Partnership}, A., et~al. 2015, \apjl, 808, L3

\bibitem[\protect\citeauthoryear{{Andrews} et~al.}{{Andrews}
  et~al.}{2018a}]{Andrews2018dsharp}
{Andrews}, S.~M., et~al. 2018a, \apjl, 869, L41

\bibitem[\protect\citeauthoryear{{Andrews} et~al.}{{Andrews}
  et~al.}{2018b}]{Andrews2018size}
{Andrews}, S.~M., et~al. 2018b, \apj, 865, 157

\bibitem[\protect\citeauthoryear{{Andrews} et~al.}{{Andrews}
  et~al.}{2011}]{Andrews2011}
{Andrews}, S.~M., et~al. 2011,
  \apj, 732, 42

\bibitem[\protect\citeauthoryear{{Andrews} et~al.}{{Andrews}
  et~al.}{2016}]{Andrews2016}
{Andrews}, S.~M., et~al. 2016, \apjl, 820, L40

\bibitem[\protect\citeauthoryear{{Ansdell} et~al.}{{Ansdell}
  et~al.}{2017}]{Ansdell2017}
{Ansdell}, M., et~al.  2017, \aj,
  153, 240

\bibitem[\protect\citeauthoryear{{Ansdell} et~al.}{{Ansdell}
  et~al.}{2018}]{Ansdell2018}
{Ansdell}, M., et~al. 2018, \apj, 859, 21

\bibitem[\protect\citeauthoryear{{Bai} \& {Stone}}{{Bai} \&
  {Stone}}{2013}]{Bai2013}
{Bai}, X.-N.,  \& {Stone}, J.~M. 2013, \apj, 769, 76

\bibitem[\protect\citeauthoryear{{Balbus} \& {Hawley}}{{Balbus} \&
  {Hawley}}{1991}]{Balbus1991}
{Balbus}, S.~A.,  \& {Hawley}, J.~F. 1991, \apj, 376, 214



\bibitem[\protect\citeauthoryear{{Birnstiel}, {Dullemond}, \&
  {Brauer}}{{Birnstiel} et~al.}{2010}]{Birnstiel2010}
{Birnstiel}, T., et~al. 2010, \aap, 513, A79

\bibitem[\protect\citeauthoryear{{Boley}}{{Boley}}{2009}]{Boley2009}
{Boley}*, A.~C. 2009, \apj, 695, 53


\bibitem[\protect\citeauthoryear{{Boley}}{{Boley}}{2017}]{Boley2017}
{Boley}*, A.~C. 2017, \apj, 850, 103

\bibitem[\protect\citeauthoryear{{Boley}, {Granados Contreras}, \&
  {Gladman}}{{Boley} et~al.}{2016}]{Boley2016}
{Boley}*, A.~C., et~al. 2016, \apjl,
  817, L17

\bibitem[\protect\citeauthoryear{{Boss}}{{Boss}}{1997}]{Boss1997}
{Boss}, A.~P. 1997, Science, 276, 1836



\bibitem[\protect\citeauthoryear{{Cowan} et~al.}{{Cowan}
  et~al.}{2015}]{Cowan2015}
{Cowan*}, N.~B., et~al. 2015, \pasp, 127, 311


\bibitem[\protect\citeauthoryear{{Crida}, {Morbidelli}, \& {Masset}}{{Crida}
  et~al.}{2006}]{Crida2006}
{Crida}, A., et~al. 2006, Icarus, 181, 587

\bibitem[\protect\citeauthoryear{{Cridland}, {Pudritz}, \& {Alessi}}{{Cridland}
  et~al.}{2016}]{Cridland2016}
{Cridland*}, A.~J., et~al.  2016, \mnras, 461, 3274

\bibitem[\protect\citeauthoryear{{Cridland}, {Pudritz}, \&
  {Birnstiel}}{{Cridland} et~al.}{2017}]{Cridland2017}
{Cridland*}, A.~J., et~al. 2017, \mnras, 465,
  3865

\bibitem[\protect\citeauthoryear{{Cridland} et~al.}{{Cridland}
  et~al.}{2017}]{Cridland2017chemistry}
{Cridland*}, A.~J., et~al. 2017, \mnras, 469, 3910

\bibitem[\protect\citeauthoryear{Dong, et al.}{2018}]{Dong2018gaps} Dong*, R., et al. 2018, ApJ, 866, 110

\bibitem[\protect\citeauthoryear{{Dong}, {Najita}, \& {Brittain}}{{Dong}
  et~al.}{2018}]{Dong2018spirals}
{Dong*}, R., et~al. 2018, \apj, 862, 103

\bibitem[\protect\citeauthoryear{{Dong} et~al.}{{Dong}
  et~al.}{2017a}]{Dong2017multi}
{Dong*}, R., et~al. 2017a, \apj, 843, 127

\bibitem[\protect\citeauthoryear{{Dong} et~al.}{{Dong}
  et~al.}{2017b}]{Dong2017}
{Dong*}, R., et~al. 2017b, \apj, 836, 201

\bibitem[\protect\citeauthoryear{{Dong} \& {Fung}}{{Dong} \&
  {Fung}}{2017}]{Dong2017gaps}
{Dong*}, R.,  \& {Fung}, J. 2017, \apj, 835, 146

\bibitem[\protect\citeauthoryear{Dong \& Fung}{2017}]{Dong2017spirals} Dong* R., Fung J., 2017, ApJ, 835, 38


\bibitem[\protect\citeauthoryear{{Dong} \& {Dawson}}{{Dong} \&
  {Dawson}}{2016}]{Dong2016}
{Dong*}, R.,  \& {Dawson}, R. 2016, \apj, 825, 77

\bibitem[\protect\citeauthoryear{{Dong} et~al.}{{Dong}
  et~al.}{2015a}]{Dong2015spirals}
{Dong*}, R., et~al. 2015a, \apjl, 809,
  L5

\bibitem[\protect\citeauthoryear{{Dong} et~al.}{2015}]{Dong2015b} Dong*, R., et al. 2015b, ApJ, 809, 93

\bibitem[\protect\citeauthoryear{Dong, et al.}{2012}]{Dong2012} Dong*, R., et al., 2012, ApJ, 750, 161

\bibitem[\protect\citeauthoryear{{Evans} et~al.}{{Evans}
  et~al.}{2009}]{Evans2009}
{Evans}, N.~J., II, et~al. 2009, \apjs, 181, 321


\bibitem[\protect\citeauthoryear{{Fedele} et~al.}{{Fedele}
  et~al.}{2010}]{Fedele2010}
{Fedele}, D., et~al. 2010, \aap, 510, A72


\bibitem[\protect\citeauthoryear{{Flaherty} et~al.}{{Flaherty}
  et~al.}{2018}]{Flaherty2018}
{Flaherty}, K.~M., et~al. 2018, \apj, 856, 117

\bibitem[\protect\citeauthoryear{{Follette} et~al.}{{Follette}
  et~al.}{2017}]{Follette2017}
{Follette}, K.~B., et~al. 2017, \aj, 153, 264

\bibitem[\protect\citeauthoryear{{Fung} \& {Lee}}{{Fung} \&
  {Lee}}{2018}]{FungLee2018}
{Fung*}, J.,  \& {Lee}, E.~J. 2018, \apj, 859, 126

\bibitem[\protect\citeauthoryear{{Fung}, {Shi}, \& {Chiang}}{{Fung}
  et~al.}{2014}]{Fung2014}
{Fung*}, J., et~al. 2014, \apj, 782, 88

\bibitem[\protect\citeauthoryear{Fung \& Dong}{2015}]{Fung2015} Fung* J., Dong R., 2015, ApJL, 815, L21

\bibitem[\protect\citeauthoryear{{Fung} \& {Chiang}}{{Fung} \&
  {Chiang}}{2017}]{FungChiang2017}
{Fung*}, J.,  \& {Chiang}, E. 2017, \apj, 839, 100

\bibitem[\protect\citeauthoryear{{Furuya} \& {Aikawa}}{{Furuya} \&
  {Aikawa}}{2014}]{Furuya2014}
{Furuya}, K.,  \& {Aikawa}, Y. 2014, \apj, 790, 97


\bibitem[\protect\citeauthoryear{{Granados Contreras} \& {Boley}}{{Granados
  Contreras} \& {Boley}}{2018}]{Granados2018}
{Granados Contreras*}, A.,  \& {Boley}, A. 2018, \aj, 155, 139

\bibitem[\protect\citeauthoryear{{Hartmann} et~al.}{{Hartmann}
  et~al.}{1998}]{Hartmann1998}
{Hartmann}, L., et~al. 1998, \apj,
  495, 385

\bibitem[\protect\citeauthoryear{{Hasegawa} \& {Pudritz}}{{Hasegawa} \&
  {Pudritz}}{2011}]{Hasegawa2011}
{Hasegawa*}, Y.,  \& {Pudritz}, R.~E. 2011, \mnras, 417, 1236

\bibitem[\protect\citeauthoryear{{Hasegawa} \& {Pudritz}}{{Hasegawa} \&
  {Pudritz}}{2013}]{Hasegawa2013}
{Hasegawa*}, Y.,  \& {Pudritz}, R.~E. 2013, \apj, 778, 78

\bibitem[\protect\citeauthoryear{{Hayashi}}{{Hayashi}}{1981}]{Hayashi1981}
{Hayashi}, C. 1981, PTPS, 70, 35

\bibitem[\protect\citeauthoryear{{Herbst} \& {van Dishoeck}}{{Herbst} \& {van
  Dishoeck}}{2009}]{Herbst2009}
{Herbst}, E.,  \& {van Dishoeck}, E.~F. 2009, \araa, 47, 427


\bibitem[\protect\citeauthoryear{{Huang} et~al.}{{Huang}
  et~al.}{2018}]{Huang2018}
{Huang}, J., et~al. 2018, \apjl, 869, L42

\bibitem[\protect\citeauthoryear{{Isella} et~al.}{{Isella}
  et~al.}{2019}]{Isella2019}
{Isella}, A., et~al. 2019 \apjl, 879, L25

\bibitem[\protect\citeauthoryear{{Isella} et~al.}{{Isella}
  et~al.}{2016}]{Isella2016}
{Isella}, A., et~al. 2016, PRL, 117, 251101

\bibitem[\protect\citeauthoryear{{Keppler} et~al.}{{Keppler}
  et~al.}{2018}]{Keppler2018}
{Keppler}, M., et~al. 2018, \aap, 617, A44

\bibitem[\protect\citeauthoryear{{Kley} \& {Nelson}}{{Kley} \&
  {Nelson}}{2012}]{KleyNelson2012}
{Kley}, W.,  \& {Nelson}, R.~P. 2012, \araa, 50, 211

\bibitem[\protect\citeauthoryear{{Lambrechts} \& {Johansen}}{{Lambrechts} \&
  {Johansen}}{2012}]{Lambrechts2012}
{Lambrechts}, M.,  \& {Johansen}, A. 2012, \aap, 544, A32


\bibitem[\protect\citeauthoryear{{Lee}}{{Lee}}{2019}]{Lee2019}
{Lee*}, E.~J. 2019, \apj, 878, 36

\bibitem[\protect\citeauthoryear{{Lee} \& {Chiang}}{{Lee} \&
  {Chiang}}{2016}]{LeeChiang2016}
{Lee*}, E.~J.,  \& {Chiang}, E. 2016, \apj, 817, 90

\bibitem[\protect\citeauthoryear{{Levison}, {Kretke}, \& {Duncan}}{{Levison}
  et~al.}{2015}]{Levison2015}
{Levison}, H.~F., et~al. 2015, \nat, 524, 322

\bibitem[\protect\citeauthoryear{{Lodato} \& {Facchini}}{{Lodato} \&
  {Facchini}}{2013}]{Lodato2013}
{Lodato}, G.,  \& {Facchini}, S. 2013, \mnras, 433, 2157

\bibitem[\protect\citeauthoryear{{Long} et~al.}{{Long} et~al.}{2017}]{Long2017}
{Long}, Z.~C., et~al. 2017, \apj, 838, 62


\bibitem[\protect\citeauthoryear{{Manara}, {Morbidelli}, \& {Guillot}}{{Manara}
  et~al.}{2018}]{Manara2018}
{Manara}, C.~F., et~al.  2018, \aap, 618, L3

\bibitem[\protect\citeauthoryear{{Manara} et~al.}{{Manara}
  et~al.}{2016}]{Manara2016}
{Manara}, C.~F., et~al. 2016, \aap, 591, L3

\bibitem[\protect\citeauthoryear{{Mann} et~al.}{{Mann} et~al.}{2014}]{Mann2014}
{Mann*}, R.~K., et~al. 2014, \apj, 784, 82

\bibitem[\protect\citeauthoryear{{Marino}, {Perez}, \& {Casassus}}{{Marino}
  et~al.}{2015}]{Marino2015}
{Marino}, S., et~al. 2015, \apjl, 798, L44

\bibitem[\protect\citeauthoryear{Marleau \& Cumming}{2014}]{marleau} Marleau* G.-D., Cumming A., 2014, MNRAS, 437, 1378

\bibitem[\protect\citeauthoryear{{Marois} et~al.}{{Marois}
  et~al.}{2008}]{Marois2008}
{Marois*}, C., et~al. 2008, Science, 322,
  1348

\bibitem[\protect\citeauthoryear{{Miotello} et~al.}{{Miotello}
  et~al.}{2016}]{Miotello2016}
{Miotello}, A., et~al. 2016,
  \aap, 594, A85

\bibitem[\protect\citeauthoryear{{Miotello} et~al.}{{Miotello}
  et~al.}{2017}]{Miotello2017}
{Miotello}, A., et~al. 2017, \aap, 599, A113

\bibitem[\protect\citeauthoryear{{Monnier} et~al.}{{Monnier}
  et~al.}{2017}]{Monnier2017}
{Monnier}, J.~D., et~al. 2017, \apj, 838, 20


\bibitem[\protect\citeauthoryear{{Muto} et~al.}{{Muto} et~al.}{2012}]{Muto2012}
{Muto}, T., et~al. 2012, \apjl, 748, L22

\bibitem[\protect\citeauthoryear{{Najita} \& {Kenyon}}{{Najita} \&
  {Kenyon}}{2014}]{Najita2014}
{Najita}, J.~R.,  \& {Kenyon}, S.~J. 2014, \mnras, 445, 3315

\bibitem[\protect\citeauthoryear{{Nielsen} et~al.}{{Nielsen}
  et~al.}{2019}]{Nielsen2019}
{Nielsen}, E.~L., et~al. 2019, ESS, 51, 6


\bibitem[\protect\citeauthoryear{{{\"O}berg} et~al.}{{{\"O}berg}
  et~al.}{2015}]{Oberg2015}
{{\"O}berg}, K.~I., et~al. 2015, \nat, 520, 198

\bibitem[\protect\citeauthoryear{{{\"O}berg}, {Murray-Clay}, \&
  {Bergin}}{{{\"O}berg} et~al.}{2011}]{Oberg2011snowline}
{{\"O}berg}, K.~I., et~al. 2011, \apjl, 743,
  L16

\bibitem[\protect\citeauthoryear{{Owen} \& {Clarke}}{{Owen} \&
  {Clarke}}{2012}]{OwenClarke2012}
{Owen*}, J.~E.,  \& {Clarke}, C.~J. 2012, \mnras, 426, L96

\bibitem[\protect\citeauthoryear{{Owen} \& {Wu}}{{Owen} \&
  {Wu}}{2013}]{OwenWu2013}
{Owen*}, J.~E.,  \& {Wu}, Y. 2013, \apj, 775, 105

\bibitem[\protect\citeauthoryear{{Owen} \& {Wu}}{{Owen} \&
  {Wu}}{2017}]{OwenWu2017}
{Owen*}, J.~E.,  \& {Wu}, Y. 2017, \apj, 847, 29

\bibitem[\protect\citeauthoryear{{Pascucci}, {Gorti}, \&
  {Hollenbach}}{{Pascucci} et~al.}{2012}]{Pascucci2012}
{Pascucci}, I., et~al. 2012, \apjl, 751, L42

\bibitem[\protect\citeauthoryear{{P{\'e}rez} et~al.}{{P{\'e}rez}
  et~al.}{2019}]{PerezS2019}
{P{\'e}rez}, S., et~al. 2019, AJ, 158, 15


\bibitem[\protect\citeauthoryear{{Pinte} et~al.}{{Pinte}  et~al.}{2019}]{Pinte2019}
{Pinte}, C., et~al. 2019, Nature Astronomy, 419

\bibitem[\protect\citeauthoryear{{Pollack} et~al.}{{Pollack}
  et~al.}{1996}]{Pollack1996}
{Pollack}, J.~B., et~al. 1996, \icarus, 124, 62


\bibitem[\protect\citeauthoryear{{Pudritz} \& {Ray}}{{Pudritz} \&  {Ray}}{2019}]{Pudritz2019} {Pudritz*}, R.~E.,  \& {Ray}, T.~P. 2019, FASS, 6, 54


\bibitem[\protect\citeauthoryear{{Rodigas} et~al.}{{Rodigas}
  et~al.}{2014}]{Rodigas2014}
{Rodigas}, T.~J., et~al. 2014, \apjl, 791, L37

\bibitem[\protect\citeauthoryear{{Rogers} \& {Wadsley}}{{Rogers} \&
  {Wadsley}}{2011}]{Rogers2011}
{Rogers*}, P.~D.,  \& {Wadsley}, J. 2011, \mnras, 414, 913

\bibitem[\protect\citeauthoryear{{Rogers} \& {Wadsley}}{{Rogers} \&
  {Wadsley}}{2012}]{Rogers2012}
{Rogers*}, P.~D.,  \& {Wadsley}, J. 2012, \mnras, 423, 1896

\bibitem[\protect\citeauthoryear{{Shakura} \& {Sunyaev}}{{Shakura} \&
  {Sunyaev}}{1973}]{ShakuraSunyaev1973}
{Shakura}, N.~I.,  \& {Sunyaev}, R.~A. 1973, \aap, 500, 33

\bibitem[\protect\citeauthoryear{{Strom} et~al.}{{Strom}
  et~al.}{1989}]{Strom1989}
{Strom}, K.~M., et~al. 1989, \aj, 97, 1451

\bibitem[\protect\citeauthoryear{{Takahashi} \& {Inutsuka}}{{Takahashi} \&
  {Inutsuka}}{2016}]{Takahashi2016}
{Takahashi}, S.~Z.,  \& {Inutsuka}, S.-i. 2016, \aj, 152, 184

\bibitem[\protect\citeauthoryear{{Tazzari} et~al.}{{Tazzari}
  et~al.}{2018}]{Tazzari2017}
{Tazzari}, M., et~al. 2018, MNRAS, 476, 4527

\bibitem[\protect\citeauthoryear{{Trapman} et~al.}{{Trapman}
  et~al.}{2019}]{Trapman2019}
{Trapman}, L., et~al. 2019, A\&A, 629, 79

\bibitem[\protect\citeauthoryear{{Tripathi} et~al.}{{Tripathi}
  et~al.}{2017}]{Tripathi2017}
{Tripathi}, A., et~al. 2017,
  \apj, 845, 44

\bibitem[\protect\citeauthoryear{{Tychoniec} et~al.}{{Tychoniec}
  et~al.}{2018}]{Tychoniec2018}
{Tychoniec}, {\L}., et~al. 2018, \apjs, 238, 19

\bibitem[\protect\citeauthoryear{{van der Marel}}{{van der
  Marel}}{2017}]{vanderMarel2017}
{van der Marel*}, N. 2017, FEDYS, 445, 39

\bibitem[\protect\citeauthoryear{{van der Marel} et~al.}{{van der Marel}
  et~al.}{2019}]{vanderMarel2019}
{van der Marel*}, N., et~al. 2019, ApJ, 872, 112

\bibitem[\protect\citeauthoryear{{van der Marel} et~al.}{{van der Marel}
  et~al.}{2015a}]{vanderMarel2015vla}
{van der Marel*}, N., et~al. 2015a, \apjl, 810, L7

\bibitem[\protect\citeauthoryear{{van der Marel} et~al.}{{van der Marel}
  et~al.}{2016}]{vanderMarel2016-isot}
{van der Marel*}, N., et~al.
  2016, \aap, 585, A58

\bibitem[\protect\citeauthoryear{{van der Marel} et~al.}{{van der Marel}
  et~al.}{2013}]{vanderMarel2013}
{van der Marel*}, N., et~al. 2013, Science, 340, 1199

\bibitem[\protect\citeauthoryear{{van der Marel} et~al.}{{van der Marel}
  et~al.}{2015b}]{vanderMarel2015-12co}
{van der Marel*}, N., et~al. 2015b, \aap, 579, A106

\bibitem[\protect\citeauthoryear{{van der Marel} et~al.}{{van der Marel}
  et~al.}{2018a}]{vanderMarel2018}
{van der Marel*}, N., et~al. 2018a, \apj, 854, 177

\bibitem[\protect\citeauthoryear{{van der Marel}, {Williams}, \&
  {Bruderer}}{{van der Marel} et~al.}{2018b}]{vanderMarel2018-hd16}
{van der Marel*}, N., et~al. 2018b, \apjl, 867,
  L14

\bibitem[\protect\citeauthoryear{{Wahhaj} et~al.}{{Wahhaj}
  et~al.}{2015}]{Wahhaj2015}
{Wahhaj}, Z., et~al. 2015, \aap, 581, A24


\bibitem[\protect\citeauthoryear{{Ward}}{{Ward}}{1997}]{Ward1997}
{Ward}, W.~R. 1997, \icarus, 126, 261

\bibitem[\protect\citeauthoryear{{Weidenschilling}}{{Weidenschilling}}{1977}]{Weidenschilling1977}
{Weidenschilling}, S.~J. 1977, \mnras, 180, 57

\bibitem[\protect\citeauthoryear{{White} \& {Boley}}{{White} \&
  {Boley}}{2018}]{White2018}
{White*}, J.~A.,  \& {Boley}, A.~C. 2018, \apj, 859, 103

\bibitem[\protect\citeauthoryear{{White} et~al.}{{White}
  et~al.}{2016}]{White2016}
{White*}, J.~A., et~al. 2016, \apj, 829, 6

\bibitem[\protect\citeauthoryear{{Williams} \& {Best}}{{Williams} \&
  {Best}}{2014}]{WilliamsBest2014}
{Williams}, J.~P.,  \& {Best}, W.~M.~J. 2014, \apj, 788, 59



\bibitem[\protect\citeauthoryear{{Winn} \& {Fabrycky}}{{Winn} \&
  {Fabrycky}}{2015}]{Winn2015}
{Winn}, J.~N.,  \& {Fabrycky}, D.~C. 2015, \araa, 53, 409

\bibitem[\protect\citeauthoryear{{Youdin} \& {Goodman}}{{Youdin} \&
  {Goodman}}{2005}]{Youdin2005}
{Youdin}, A.~N.,  \& {Goodman}, J. 2005, \apj, 620, 459

\bibitem[\protect\citeauthoryear{{Zhang}, {Blake}, \& {Bergin}}{{Zhang}
  et~al.}{2015}]{Zhang2015}
{Zhang}, K., et~al. 2015, \apjl, 806, L7

\bibitem[\protect\citeauthoryear{{Zhang} et~al.}{{Zhang}
  et~al.}{2018}]{Zhang2018dsharp}
{Zhang}, S., et~al. 2018, \apjl, 869, L47

\end{thebibliography}
\end{document}